\RequirePackage{lineno}
\documentclass[prl, etal, twocolumn,  longbibliography, amsmath,superscriptaddress,amssymb]{revtex4-1}
\usepackage{graphicx}% Include figure files
\usepackage{subfigure}
\usepackage{dcolumn}% Align table columns on decimal point
\usepackage{bm}% bold math
\usepackage{verbatim}
\usepackage{xcolor}
\usepackage{braket}
\usepackage{tikz}
\usetikzlibrary{shapes.geometric, arrows}
\usepackage{pgfplots}
\usepackage{mathtools}
\usepackage{color}

\begin{document}

%\linenumbers

%\preprint{}
%\begin{CJK*}{UTF8}{gbsn}
\title{Femtosecond pumping of nuclear isomeric states by  
 the Coulomb  collision of ions with quivering electrons}

\def\fdu {Key Laboratory of Nuclear Physics and Ion-beam Application (MoE), Institute of Modern Physics, Fudan University, Shanghai 200433,  China}
\def\sjtu {Key Laboratory of Laser Plasma (MoE), School of Physics and Astronomy, Shanghai Jiao Tong University, Shanghai 200240, China}
\def\ifsa {IFSA Collaborative Innovation Center, Shanghai Jiao Tong University, Shanghai 200240,  China}
\def\SARI{Shanghai  Advanced Research Institute, Chinese Academy of Sciences, Shanghai 201210,  China}
\def\LOP{Laboratory of Optical Physics, Institute of Physics, Chinese Academy of Sciences, Beijing 100190, China}
\def\UCA{School of Physical Sciences, University of Chinese Academy of Sciences, Beijing 100049,  China}

\author{Jie Feng}      \affiliation{\sjtu}
\author{Wenzhao Wang} \affiliation{\sjtu}
\author{Changbo Fu} \email[Corresponding author: ] {cbfu@fudan.edu.cn} \affiliation{\fdu}
\author{Liming Chen} \email[Corresponding author: ] {lmchen@sjtu.edu.cn}\affiliation{\sjtu}\affiliation{\ifsa}
\author{Junhao Tan} 	\affiliation{\sjtu}
\author{Yaojun Li} 	\affiliation{\sjtu}
\author{Jinguang Wang} \affiliation{\LOP}
\author{Yifei Li} 	\affiliation{\LOP}
\author{Guoqiang Zhang}  \affiliation{\SARI}
\author{Yugang Ma} \affiliation{\fdu}
\author{Jie Zhang} \affiliation{\sjtu}\affiliation{\ifsa}

%\altaffiliation[also at]{}
%Lines break automatically or can be forced with \\

\date{\today}% It is always \today, today,
             %  but any date may be explicitly specified
%=================================================================
%=================================================================
%\pacs{25.70.Ef, 29.27.-a}% PACS, the Physics and Astronomy
                             % Classification Scheme.

\begin{abstract}

Efficient production of nuclear isomers is critical for pioneering applications, like nuclear clocks, nuclear batteries, clean nuclear energy, and nuclear gamma-ray lasers. However, due to small production cross sections and quick decays, it is extremely difficult to acquire significant amount of isomers with short lifetimes via traditional accelerators or reactors  because of low beam intensity. Here, for the first time, we experimentally present femtosecond pumping of nuclear isomeric states by the Coulomb excitation of ions with the quivering electrons induced by laser fields.  Nuclei populated on the third excited state of $^{83}$Kr are generated with a peak efficiency of $2.34\times 10^{15}$ particles/s  from a table-top hundred-TW laser system.  It can be explained by the Coulomb excitation of ions with the quivering electrons during the interaction between laser pulses and clusters at nearly solid densities. This efficient and universal production method can be widely used for pumping isotopes with excited state lifetimes down to ps, and could benefit for fields like  nuclear transition mechanisms, nuclear gamma-ray lasers.
\end{abstract}
\maketitle

%\tableofcontents
%=================================================================
\maketitle

%==============================================
%============================

%\section{\uppercase\expandafter{\romannumeral1}. INTRODUCTION}
Nuclear isomers have a broad range of applications\cite{Hf178-carroll2002initial,Ta80-belic1999photoactivation, X-Dr-Gama-carroll2001x, Mo93-gunst2014dominant, NEEC-wu2019Mo93m, Tc99m-banerjee2001evolution, CExcit-RMP-alder1956study, SungYang2005-NPhys}. For example, nuclear isomers like $^{178m2}$Hf and $^{180m}$Ta etc., are regarded to be very good battery materials\cite{Hf178-carroll2002initial,Ta80-belic1999photoactivation, X-Dr-Gama-carroll2001x, Mo93-gunst2014dominant, NEEC-wu2019Mo93m}, due to their extremely high energy storage capabilities compared with chemical ones; $^{99m}$Tc isomers are widely used in medical radiographic imaging\cite{Tc99m-banerjee2001evolution}; nuclear isomers like $^{229m}$Th etc. are possible candidates for the next generation of atomic clocks as the most accurate time and frequency standards\cite{Th229-clock-von2016direct, WangXuPRL}; nuclear gamma-ray lasers were also proposed\cite{NGL-tkalya2011proposal}. Nuclear isomers also play important roles in nucleosynthesis\cite{Waker.Dracoulis_2016isomer,IsomersinAstrophysics}, that is relevant with the creation of the nuclear isotopes in stars, and then eventually affect the creation of life in the cosmos.

For these potential applications shown above, especially for nuclear gamma-ray lasers, where the lifetimes of excited states as short as nanosecond or even shorter are required\cite{GLaser-rivlin2010nuclear, GLaser-LPR}, due to the relatively large emission linewidth of Doppler broadening.
 From the experimental point of view, the bottleneck  of nuclear gamma-ray lasers appears as how to  pump excited states efficiently. Traditionally, isomers are produced with accelerators or reactors. However, limited by the beam intensities of these drivers, it is very difficult to accumulate enough amounts of isomers, in many cases, especially for those  short-lived isomers  or extremely unstable excited states.

The rapid development of high-intensity femtosecond laser brings great potentials to new concept of accelerators and radiation sources\cite{Nucl.Laser.GOBET201180,
Laser-Acc-tajima1979laser,ditmire1997high,Remington1999,DDn-Nature1999,X-ray-2003Issac,Fe57-2011Golovin}. Nowadays, the laser intensity focused onto targets can be beyond 10$^{22}$ W/cm$^2$. It can create hot plasmas with extreme pressure, temperature, and currents under which the nuclear reactions can take place\cite{DDn-Nature1999,Laser-Collider-fu2015laser,Exp.Low.Ex.fs.2014LaserPhys}. Here we report the first proof of principle experiment of femtosecond pumping of $^{83}$Kr to its  excited states by the Coulomb excitation of quivering electrons collision with ions during laser-cluster interactions. A significant amount of isomers has been detected. This opens a new path to produce nuclear isomers with an extremely high efficiency during an extremely short time. This abnormally high efficiency may lead to a deeper understanding of the laser-isomeric quantum physics.

%============================
%\section{\uppercase\expandafter{\romannumeral2}. EXPERIMENTAL SETUP}
\begin{figure*}
 \centering
\includegraphics[width=16cm]{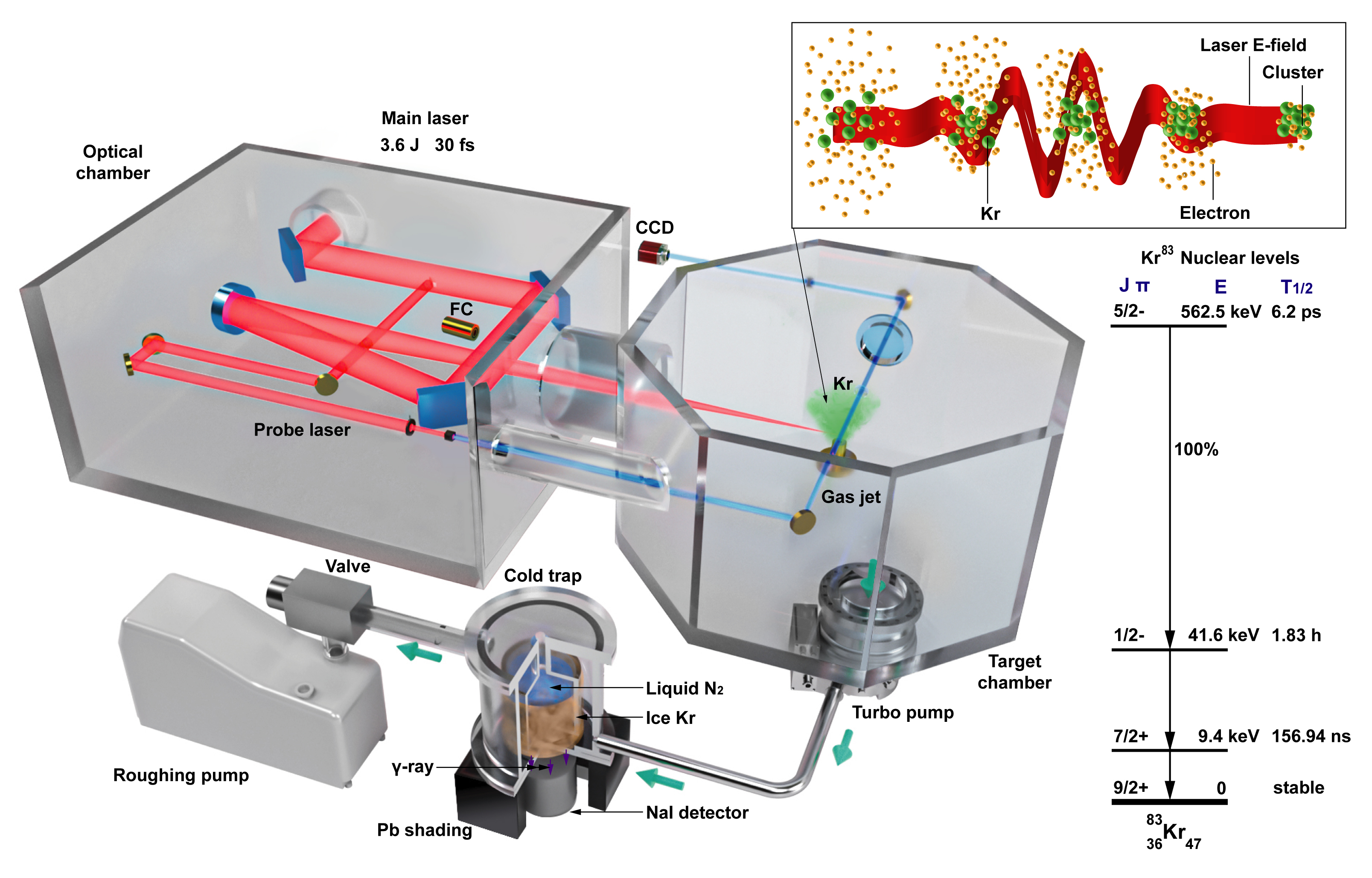}
 \caption{The schematic experimental setup. The femtosecond laser pulses are focused on the Kr nanoparticles ejected from a high-pressure Kr gas nozzle\cite{Boldarev2017RN68,ZhangLu}. A Faraday cup (FC) is used to measure the ion spectrum. A probe laser pulse is aligned through the gas/plasma during interaction to provide interferograms\cite{Abel-inv-Benattar1979}. After the interaction between the laser and clusters, the gas is pumped and frozen on surfaces of the liquid-nitrogen cold trap. The possible radioactive decays are recorded by a NaI detector under the trap. { The right inset shows} the decay scheme of $^{83}$Kr. The J$\pi$, energies, and half lifetimes of the ground state (0), { the} first (9.4 keV), second (41.6 keV) and third (562.5 keV) excited states are listed.
}
\end{figure*}
\paragraph{Experimental setup.}
The experiment is carried out using a Ti: Sapphire laser system at the Laboratory of Laser Plasmas of Shanghai Jiao Tong University. The experimental setup is shown in Fig. 1. { Laser} pulses with an energy of 3.6 J { and duration of 30 fs (FWHM)} are focused at an intensity of $1\times$10$^{19}$ W/cm$^2$ on natural Kr gas, with 11.5\% $^{83}$Kr isotope, from a jet in the vacuum chamber with a backing pressure up to 7 MPa. During the adiabatic expansion process, the Kr gas { jet is cooled} down and { frozen into} nanoparticles or clusters which { serves} as the { targets}. The radius of the clusters have been measured with Rayleigh scattering approach\cite{RN71,RN72}. During the laser-plasma interaction, Kr clusters are ionized, and then electrons are rapidly heated by various collective and nonlinear processes to a non-equilibrium state with mean energies of many keV before the clusters disassemble in the laser field\cite{ditmire1997high}. During this process, the low-lying excited states of $^{83}$Kr, including the first excited isomeric state $^{83m}$Kr$_1$ with $E_1$ = 9.4 keV, the second $^{83m}$Kr$_2$ with $E_2$ = 41.6 keV and the third $^{83m}$Kr$_3$ with $E_3$ = 562.5 keV, could be populated due to physical processes to be discussed later. After the laser-cluster interaction, the Kr gas is collected by a turbo pump, and then goes into a cold trap to be frozen on and reserved for the subsequent detection.

To improve the gas collecting efficiency, vacuum pumps in the system are used as follows. There are two pump sets on the optical chamber and target chamber respectively. Each set has a roughing pump and a turbo pump. Before the experiment, they are turned on to keep the { high} vacuum. During experiment, the pump set at the optical chamber, as well as the roughing pump at the target chamber, are turned off. Only the turbo pump at the target chamber is on. In this way, we estimate that over 95\% Kr gas can be collected onto the cold trap. The cold trap is cooling down with liquid nitrogen. There is a 380 $\mu$m thick Be window close to the trap. If Kr isomers are produced during experiment, $\gamma$-rays from the isomer decays would be detected later by a NaI detector located behind the window. The NaI detector is composed of a NaI crystal of 3 mm thickness and a Be window of 200 $\mu$m thickness. Because of { relatively poor energy resolution of the NaI detector}, individual peaks { of} $^{55}$Fe(5.9 keV),$^{241}$Am(59.5 keV), $K_\alpha$ fluorescence Cu(8.0 keV) and Mo(17.5 keV) are used to { calibrate the detector$'$s energy resolution and efficiency}.

The cold-trap detecting efficiency is the { result} of the following factors: the chamber-to-trap gas transferring efficiency (95$\%$), the area ratio of the cold trap$'$s bottom to the whole surface (26.5$\%$), the factor due to Be window absorption (95.6$\%$), the factor due to the delayed measuring time (81.8$\%$), the geometrical factor due to the distance between the detector and the Kr ice (4.7$\%$), and the decay branch ratio (21.39$\%$). Therefore, we estimate the cold-trap detecting efficiency to be 0.20$\%$ in X-ray energy range of 9 to 15 keV. The experimental signals, as well as the backgrounds were measured in this way: Before the laser shooting, turned the NaI detector on, and took data for 4 hours, which was used as background. Then turn the NaI detector off, and shoot the gas target by laser for about 1 hour { at repetition rate of 0.025 Hz}, with the liquid nitrogen in the cold-trap for collecting Kr gas. After that, turn off the laser, and after about 3 mins, turn the NaI detector on, and take nuclear radiative data for about 4 hours.

%============================
%\section{\uppercase\expandafter{\romannumeral3}. RESULTS AND ANALYSIS}
\paragraph{Ions and temperature.}
A typical time-of-flight spectrum obtained by the Faraday cup is shown in the supplementary. The energy spectrum of Kr ions from the coulomb explosions follows the quasi-Boltzmann distribution\cite{Barbui.PRL.DD, Barbui.PRL.DD1}. The temperature of Kr ions is fitted to be { $T= (15\pm 4)$} keV. By integrating the area under the quasi-Boltzmann distribution curve, and taking the average Kr charge state to be 14$^+$ according to ADK model\cite{Charge-state-delone1986tunnel}, we estimate the total number of the Kr atoms in the laser spot to be 4.5$\times$10$^{15}$ in one shot, and then the total number of the $^{83}$Kr atoms is
\begin{equation}
N_t^{FC}=(5.0\pm 0.3)\times 10^{14} {\rm/shot},
\end{equation}
where the error comes from the fitting of the Faraday cup spectrum.

\begin{figure}[b]
 \centering
	\includegraphics[width=0.45\textwidth]{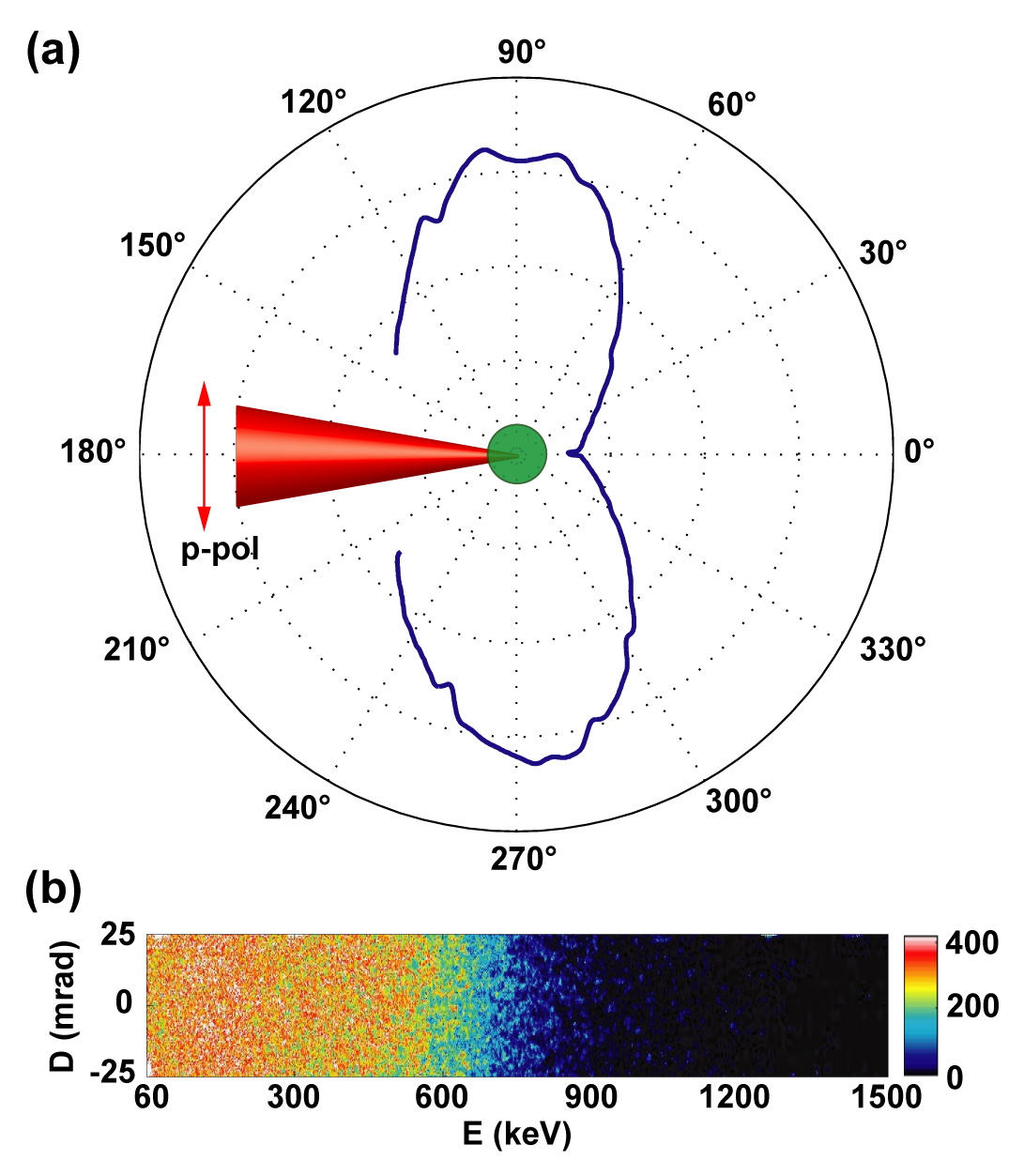}
 \caption{Experimental electron beam results. (a) Electron beam angular distribution. The laser pulses with p-polarization propagate at $0^\circ$ direction. Electron beam are recorded on an Image Plate which is covered by a 15 $\mu$m Aluminum foil. (b) Electron spectrum measured at $90^{\circ}$ direction for twenty cumulative shots and recorded on Image Plate.}
%\label{fig.e.spec}
\end{figure}

\paragraph{Electron angular distribution and energy.}
The electrons angular distribution was measured by a curved image plate (IP). The IP covered $280^\circ$ horizontally and $50^\circ$ vertically. The electrons spectra were measured by a magnetic spectrometer located at $90^\circ$ to the laser beam direction with magnetic field 0.15 T, and the minimum detection energy is about 60 keV. The electrons angular distribution is shown in Fig. 2(a). Two electron beams appear in the direction of laser polarization which is the typical phenomenon of laser-cluster resonance. When the laser intensity  normalized vector potential\cite{Eric_RMP_a0} $a_0 <1$, the cluster electrons mainly oscillate along the laser polarization direction which is called linear resonant mechanism\cite{cluster-PRL.92.133401, Ditmire1996RN64}. However, when the laser intensity $a_0 >1$, these electrons would oscillate like the shape of number eight and acquire energy from laser electric-field quickly up to hundreds of keV { or} MeV, which is called non-linear resonant mechanism\cite{kundu2006nonlinear,Ar-cluster-PhysRevLett.104.215004}. As shown in Fig. 2(b), electron energy distribution is between hundreds of keV to MeV which is higher enough to { pump} $^{83}$Kr to its excited states.

\paragraph{$^{83m}$Kr$_2$ { generation} efficiency.}
A typical energy spectrum measured by the NaI detector is shown in Fig. 3(a). The decay spectrum is shown in Fig. 3(b). As we know from the NNDC database\cite{BNL-web}, for isomeric state of $^{83m}$Kr$_2$ which half life is 1.83 h, there are several decay lines, i.e. 12.6 { keV} ($K_\alpha$), 14.1 { keV}($K_\beta$) (internal conversion decay from $^{83m}$Kr$_2$ to $^{83m}$Kr$_1$) and 9.4 keV (from $^{83m}$Kr$_1$ to ground state) with { branch ratios (B.R.)} of 13.8\%, 2.1\% and 5.5\%, respectively. In our experiment, the half life is measured to be 1.80$\pm$0.05 h, agrees well with $^{83m}$Kr$_2$ half-life. Both the energy and time spectra prove that the decays are from $^{83m}$Kr$_2$. The total number of radiation photons detected from $^{83m}$Kr$_2$ isomers { for} 100 shots is fitted to be 2283$\pm$30 particles. For each experimental run, we have 100 shots with the same shooting speed in about 67 mins. Considering each shot with an energy of 3.6 J, as well as the cold-trap detecting efficient of 0.20\%, we deduce that the $^{83m}$Kr$_2$ producing efficiency for single shot is
\begin{equation}\label{eq.N2.exp}
N_2^{exp}=(1.15\pm 0.02)\times 10^4 {\rm p/shot}=(3203\pm42){\rm p/J}.
\end{equation}
{ Considering the total yield of the Kr atoms in the laser spot and the abundance of $^{83}$Kr is 11.5\%, the reaction ratio from $^{83}$Kr ground state to $^{83m}$Kr$_2$ isomeric state is estimated to be $2.2\times 10^{-11}$.}
\begin{figure}[!h]
 \centering
	\includegraphics[width=0.45\textwidth]{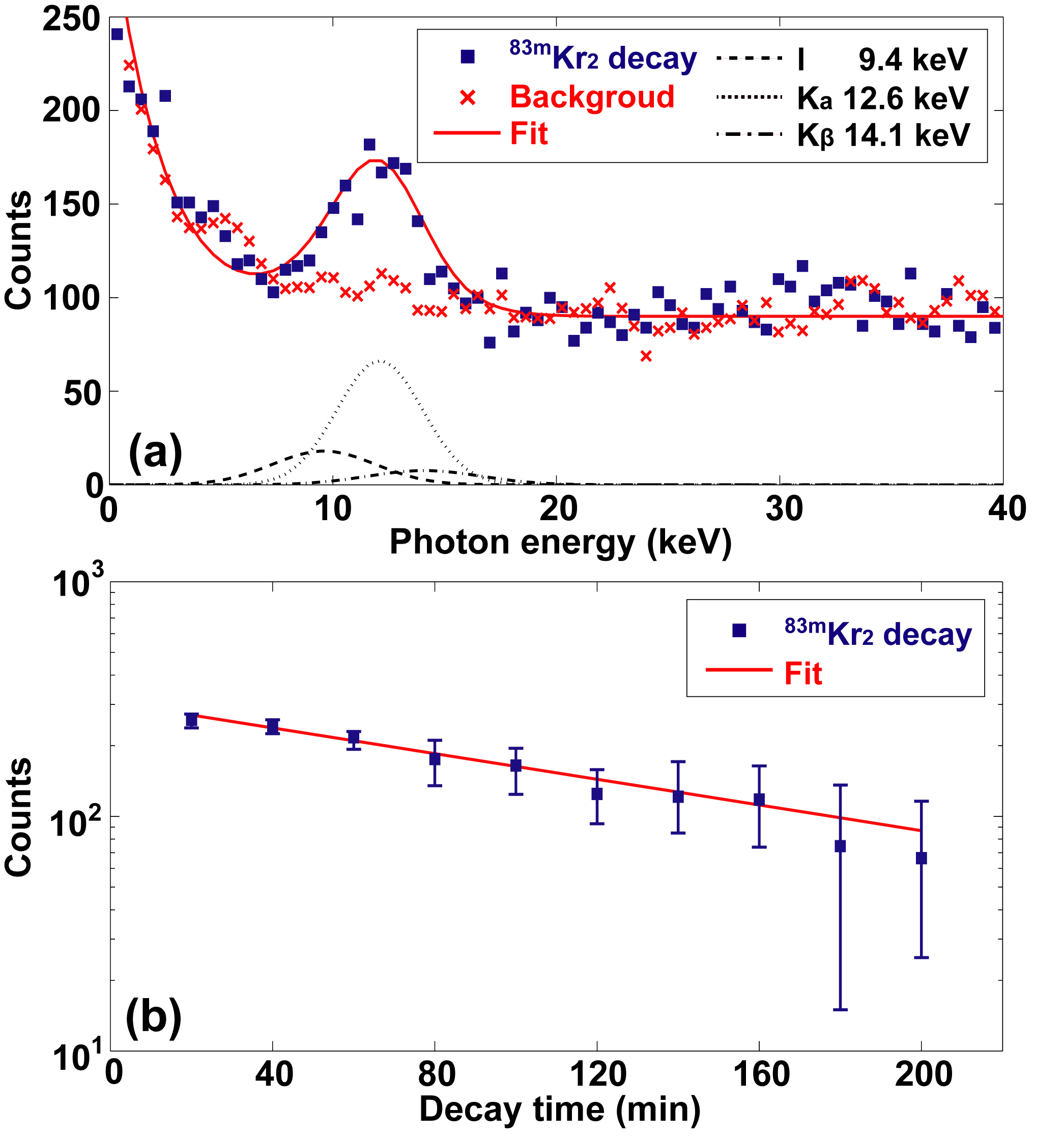}
 \caption{{Detected} $^{83}$Kr isomeric state and decay characteristics. (a) A typical { energy} spectrum measured by the NaI detector. The blue square markers represent the data measured while the red cross the background. 
 The red solid line is the fit with sum of the decay (9.4 keV, { B.R.=5.5\%}, black dash line), $K_\alpha$ (12.6 keV, { B.R.=13.8\%}, black dot line), and $K_\beta$ (14.1 keV, { B.R.=2.1\%}, black dot-dash line), according to the NNDC database\cite{BNL-web}. 
 (b) The decay time spectrum. Blue squares represent the average values of three sets of experimental data, and error bar denotes the { maximum and minimum values}. The red solid line represents a fit with an exponential function. The fitted half-life is $T_{1/2}^{exp}=(1.80\pm 0.05)$ h.}
%\label{fig.e.spec}
\end{figure}

\begin{figure*}[t]
 \centering
\includegraphics[width=16.8cm]{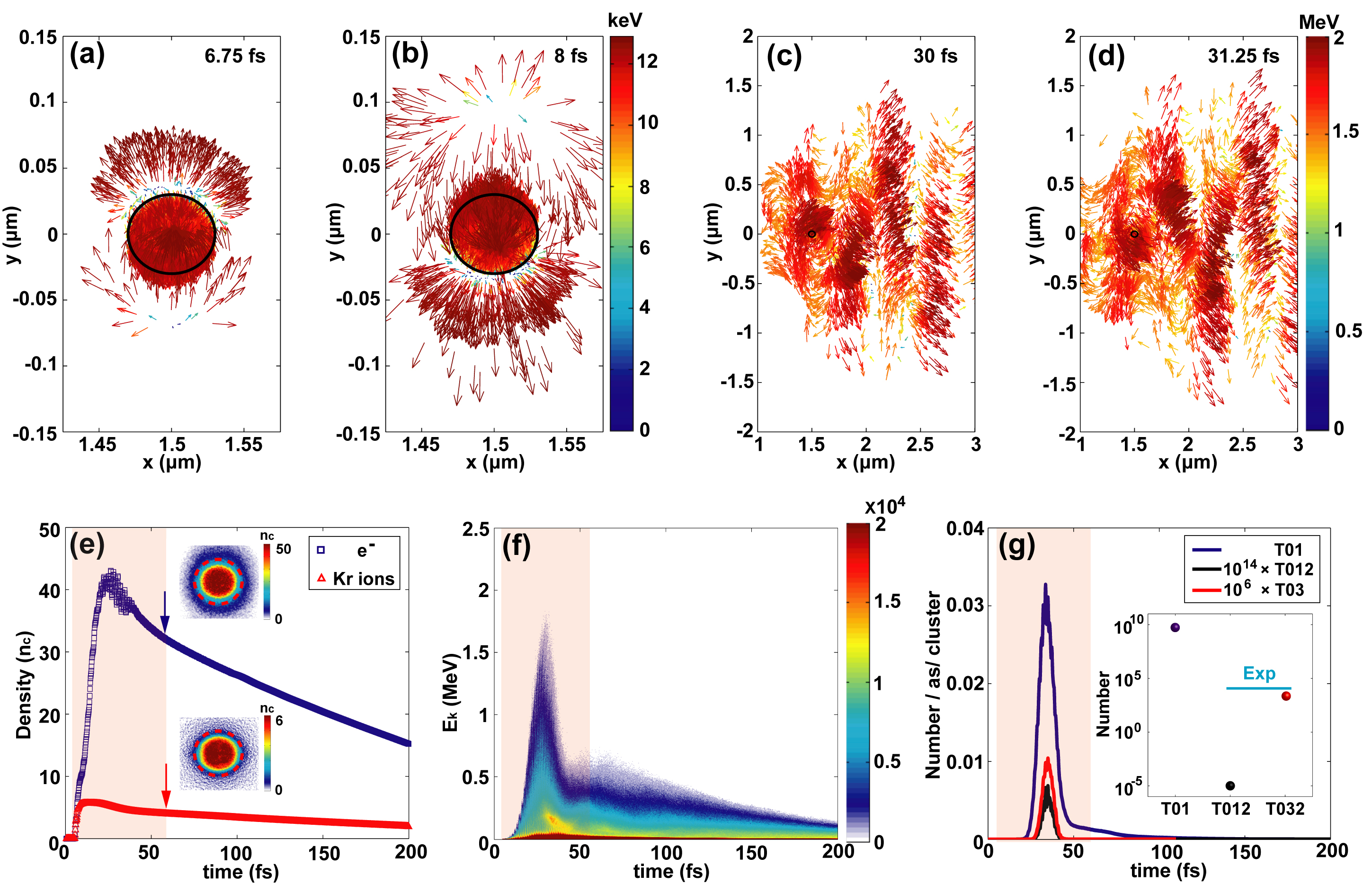}
\caption{Particle-in-cell simulation of nuclear Coulomb
excitation in interaction of laser pulses with cluster. (a-d)
Laser driven electrons from a cluster { at} four
different { times} (t = 6.75 fs, 8 fs and t = 30 fs, 31.25 fs
correspond to the process of linear resonance\cite{cluster-PRL.92.133401, Ditmire1996RN64} and nonlinear resonance\cite{kundu2006nonlinear,Ar-cluster-PhysRevLett.104.215004} respectively), where the arrow
represents electron$'$s motion direction and the color shows
electron$'$s energy. (e) The average density evolutions of electron
and krypton ions in cluster region, and the insets are the
density distributions at t = 60 fs, the $n_c$ is critical
density equaling to $1.74\times 10^{21}$ cm$^{-3}$, { the red dashed cycles in the insets represent} the original cluster region. (f) The energy evolution of electrons in cluster region. (g) The excitation number evolution for three different energy states, and the inset is total excitation number in the laser irradiation region. The light red shaded { areas in Figs. (e-g)} represent the temporal window of laser-cluster interaction.}
\end{figure*}

\paragraph{Possible mechanisms of $^{83m}$Kr$_2$ { generation}.}
Many processes could, of course, be responsible for the production of $^{83}$Kr isomers in our { experiment},
\begin{equation}
A+ {^{83}{\rm Kr}} \rightarrow {^{83m}{\rm Kr}_i+A},
\end{equation}
where $i$ =1, 2, or 3…, representing different nuclear excited states of Kr. Here $A$ can be electrons, or isotopes of Kr, i.e. $^{78,80,82,83,84,86}$Kr, considering the fact that the Kr gas are natural in this experiment. Due to the fact that electron { density}, as well { as} their { energy}, are much higher than those of ions, the { contribution} from ion-ion { collision} could be neglected.

There are several possible transitions for $^{83}$Kr excitations, as shown in Fig. 1. The transitions which are responsible for the $^{83m}$Kr$_2$ could be { $g.s. \to 2nd$ (T02), $g.s. \to 1st. \to 2nd$ (T012), and $g.s. \to 3rd. \to 2nd$ (T032).} Excited levels above 3rd are also considered but having negligible contributions.

If the transitions T02, T012, and T032 are due to the Coulomb excitation mechanism, their strength could be estimated following the Ref.\cite{CExcit-RMP-alder1956study}. For electric excitation, the cross section can be written as
\begin{equation}\
\sigma_{E\lambda}=c_{E\lambda} E^{\lambda- 2}(E-\Delta E')^{\lambda-1} B(E\lambda) f_{E\lambda},
\end{equation}
where $E$ is the projectile$'$s energy, $\Delta E'=(1 +A_1/A_2)·\Delta E$ {and} $\Delta E$ represents the excitation energy, $A_1$($A_2$) is the mass of the projectile (target),$\lambda$ is the order of electric multipole component, $B$($E\lambda$) represents the reduced transition probability associated with a radiative transition of multipole order $E\lambda$, $f_{E\lambda}$ is the f-function described in Ref.\cite{CExcit-RMP-alder1956study}, $c_{E\lambda}$ is
\begin{equation}
c_{E\lambda}= \frac{Z_1^2 A_1}{40.03}
\left[0.07199(1+A_1/A_2)Z_1Z_2\right]^{-2\lambda+2},
\end{equation}
where $Z_1$($Z_2$) is the charge of the projectile (target).

The isomer { generation} through the Coulomb excitations by electrons can be estimated by
\begin{equation}\label{eq.NCE}
N^{CE}=\iint n_e n_0 \braket{\sigma_{E\lambda} v_e} {\rm d} V {\rm d} t.
\end{equation}
For magnetic excitations, the similar formula can also be established\cite{CExcit-RMP-alder1956study}. Through a Particle-in-cell (PIC) simulation\cite{Arber2015RN69}, the numbers of $^{83m}$Kr$_2$ produced during the laser cluster interaction through T02, T012, and T032 are estimated (see Fig. 4(g) and also supplementary for details).

The PIC simulation shows that the electron quiver energy has the order of MeV ({ see} Fig. 4(f)) at a laser peak intensity of $1\times 10^{19}$ W/cm$^2$, the same as that in our { experiment} (see Fig. 2(b)). The electron energy is high enough to excite $^{83}$Kr nuclei. During the laser-cluster interaction, the direction of the { energetic electrons} flips as laser fields about 20 times, which is defined as nonlinear resonant mechanism in fs laser-cluster interactions\cite{kundu2006nonlinear,Ar-cluster-PhysRevLett.104.215004}. Because of the flips, many energetic electrons can go back and forth, and collide the relative static heavy ions (Figs. 4(a)-4(d)), and then get $^{83}$Kr nuclei excited. The high densities of ions (about 5$n_c$) and electrons (about 40$n_c$) (Fig. 4(e)), as well as the high electron temperature { (Fig. 4(f))}, result in the high productivity of the { excited nuclei}. Figure. 4(g) shows the { calculated} Coulomb excitation rates for three different paths at different time, { and the comparison of total excitation number with experimental data}. For all paths, the generation rates of $^{83}$Kr { excited} states dominated { in period of laser-on}, which is about { 10 fs}. The ratio of { T01: T012: T03$\approx 5\times 10^{14}$: 1: $2\times 10^8$}. Because { the transition probability of T32 is equal to 73.8\%}, we have { T032= T03$\times$T32}. Our simulation results demonstrate that the most possible path of $^{83m}$Kr$_2$ observed in experiment is T032, which is clearly shown in the inset of Fig. 4(g). The disagreement in Fig. 4(g) { that the theoretic calculation is about five times lower than the experimental result,} may come from the PIC simulation errors, as well as other processes not considered in the previous calculations. For example, processes including nuclear excitation by electron transfer (NEET), electron bridge (EB), nuclear excitation by electron capture (NEEC), may also contribute significantly\cite{NEET-NEEC-morel2005nuclear, gunst2018NEEC,NEEC-PRA2019WuYB,Coul.exc.Plasma.2013}. { According to the theoretic model and experimental detection result, the $3^{rd}$ excited state ($E$ = 562.5 keV) has been generated with peak efficiency (p.e.) of $2.34\times10^{15}$ particles/s, { or a.e.=$390$ p/s}, and it quickly decays to the $2^{nd}$ isomeric state ($E$ = 41.6 keV) in a lower efficiency of p.e.$=5.07\times10^{14}$ p/s { or a.e.$=290$ p/s} (look supplementary for details)} Furthermore, { the} extremely short pumping period means that this method can pump any { nuclear excited states} which have lifetimes { even} down to { ps}.

%============================
%\section{\uppercase\expandafter{\romannumeral4}. DISCUSSION}
In the application of nuclear gamma-ray lasers, it is a serious challenge for traditional methods to pump nuclei to excited states efficiently in very short temporal duration. For example, considering a conventional commercial electron accelerator ($E$ = 5 MeV, $I$ = 2 mA, 500 Hz and duration of 15 $\mu$s) shooting on the same Kr clusters target, the estimated yield of $^{83m}$Kr$_2$ is about 72 for a single shot, and its p.e.$=4.8\times 10^{6}$p/s { (or a.e.=$3.6\times 10^4$ p/s)}. And also, for nuclei excitation from photonuclear reactions ($\gamma$,$\gamma'$) and ($\gamma,n$) in solid target driven by a traditional high energy electron accelerator (e.g., BEPC-\uppercase\expandafter{\romannumeral2} in China), 
{ p.e.=$10^{10}$ p/s (or a.e.=$10^6$ p/s) could be achieved.}
But, it is still impossible for these accelerators to pump and accumulate excited state nuclei with so short lifetimes.
%, due to the pump rate is much slower than the nuclei decay rate.
Our experimental result shows an efficient way to quickly pump { nuclei} in femtosecond temporal duration via femtosecond laser-cluster interaction. The method itself is universal and easy to be realized. { By forming} nano particles from any material of gas, liquid or solid, one can shot them by laser pulses, and get them to excited states.

%============================
%\section{\uppercase\expandafter{\romannumeral5}. SUMMARY}
In summary, for the first time, we have presented efficient femtosecond pumping of nuclear isomeric states by Coulomb collision of ions with quivering electrons { during laser-cluster interaction}. By { irradiation} of Kr cluster targets with a 30 fs laser pulses at 120 TW,  the $2^{nd}$ isomeric state ($E$ = 41.6 keV) and the $3^{rd}$ excited state ($E$ = 562.5 keV) have been generated with peak efficiency of $5.07\times$10$^{14}$ and $2.34\times$10$^{15}$ p/s respectively. 
 Both efficiencies are much higher than traditional methods. Our simulation shows that the high efficiency comes from the collisions of high  density ions with  extremely high density energetic electrons accelerated by the  nonlinear resonant during the laser pulses duration. 
%We also find that the $^{83m}$Kr$_2$ ($E$ = 41.6 keV) isomers are mainly produced through the middle state, i.e., { $g.s. \to 3rd \rightarrow  2nd$, and the direct transition $g.s. \rightarrow 2nd$} can be negligible. 
This high efficiency, femtosecond duration, and easy accessibility of production of short lifetime of nuclear isomers could be greatly beneficial for  the study of nuclear transition mechanisms and nuclear gamma-ray lasers.

%\section{ACKNOWLEDGEMENTS}
This work is supported by the Science Challenge Project (TZ2018005), 
the National Natural Science Foundation of China (11875191, 11991073, 11890710, 11721404), 
the Strategic Priority Research Program of the CAS (XDB1602, XDB16), 
the National Key R\&D Program of China (2017YFA0403301), and the Key Program of CAS (XDA01020304, XDB17030500). 
%We would like to acknowledge the 200 TW laser operating staffs for running the laser facility.

%\bibliographystyle{apsrev4-1}
%\bibliography{Kr83ref}
%merlin.mbs apsrev4-1.bst 2010-07-25 4.21a (PWD, AO, DPC) hacked
%Control: key (0)
%Control: author (72) initials jnrlst
%Control: editor formatted (1) identically to author
%Control: production of article title (-1) disabled
%Control: page (0) single
%Control: year (1) truncated
%Control: production of eprint (0) enabled
%

\end{document}